# When Will AI Transform Society? Swedish Public Predictions on AI Development Timelines

**Filip Fors Connolly, Mikael Hjerm & Sara Kalucza**

Department of Sociology

Umeå University

**Abstract**

This study investigates public expectations regarding the likelihood and timing of major artificial intelligence (AI) developments among Swedes. Through a mixed-mode survey (web/paper) of 1,026 respondents, we examined expectations across six key scenarios: medical breakthroughs, mass unemployment, democratic deterioration, living standard improvements, artificial general intelligence (AGI), and uncontrollable superintelligent AI. Findings reveal strong consensus on AI-driven medical breakthroughs (82.6%), while expectations for other major developments are significantly lower, ranging from 40.9% for mass unemployment down to 28.4% for AGI. Timeline expectations varied significantly, with major medical advances anticipated within 6-10 years, while more transformative developments like AGI were projected beyond 20 years. Latent class analysis identified three distinct groups: optimists (46.7%), ambivalents (42.2%), and skeptics (11.2%). The optimist group showed higher levels of self-rated AI knowledge and education, while gender differences were also observed across classes. The study addresses a critical gap in understanding temporal expectations of AI development among the general public, offering insights for policymakers and stakeholders.

**Introduction**
Artificial intelligence (AI) development has accelerated dramatically, resulting in innovations with unprecedented capabilities across various domains. Prominent examples of such breakthroughs include sophisticated language processing (Brown et al., 2020) and accurate protein structure prediction (Jumper et al., 2021). AI is now being applied in diverse societal areas, such as combating climate change (Verendel, 2023), transforming marketing (Kshetri et al., 2024), and enhancing medical imaging (Panayides et al., 2020). The significance of AI-advances was recently underscored as the scientific establishment acknowledged their transformative potential, with Geoffrey Hinton and Demis Hassabis being awarded the 2024 Nobel Prizes in Physics and Chemistry respectively for their pioneering contributions to deep learning and artificial intelligence (Nobel Prize Outreach AB., 2024).

However, this rapid progress has sparked intense debate within the scientific community about both the promises and perils of AI development. Leading researchers present starkly different visions of AI's future impact. While some emphasize its transformative potential for addressing global challenges in healthcare, scientific discovery, and economic productivity (Chui et al. 2023; Viswa et al. 2024), others warn of existential risks. Nobel laureate Hinton, often called the "godfather of AI," left his position at Google to more freely express his concerns about AI's potential dangers (Schechner & Seetharaman, 2023). Meanwhile, philosophers like Nick Bostrom (2017; 2024) have articulated both the tremendous benefits and risks of advanced AI systems, particularly emphasizing the importance of ensuring these systems remain aligned with human values and interests.

While expert opinions on AI's trajectory vary dramatically, it is also important to understand how the general public perceives these technologies. There has been some research on public attitudes toward AI (Cave et al., 2019; Selwyn & Cordoba, 2022). For example, research shows that people in the U.K are mainly anxious about the development (Cave et al., 2019) and that a large proportion of people in many countries believe that AI will have an extensive transformative effect on society (Seth, 2024). Building on this foundation, the present study investigates public perceptions of AI specifically within Sweden. Our study contributes to this growing body of research in three important ways. First, we provide a comprehensive analysis of public perceptions concerning AI's potential impacts across multiple domains. Rather than focusing solely on general attitudes, we examine specific scenarios ranging from medical breakthroughs to existential risks, providing both descriptive analyses and identifying distinct belief clusters through latent class analysis. This approach reveals important nuances in how different segments of the population conceptualize AI's future trajectory.

Second, we investigate the temporal dimensions of public expectations regarding AI development. While existing research has documented general attitudes toward AI, the question of when people expect various AI-driven changes to materialize has received little attention. The timeline for achieving transformative AI capabilities remains highly contested even among experts, with recent surveys of AI researchers revealing substantial disagreement about when we might achieve artificial general intelligence (AGI) or other major AI capabilities (c.f. Grace et al., 2024). The median prediction for AGI achievement on Metaculus, a forecasting platform, currently stands at 2031—a surprisingly near-term estimate that contrasts with more conservative academic predictions (Metaculus, 2025). Understanding these temporal expectations is crucial for several reasons. It provides essential information for policymakers navigating the complex landscape of AI governance. If experts identify significant risks from advanced AI systems but the public expects such systems to arrive much later (or vice versa), this misalignment could complicate efforts to implement appropriate regulatory frameworks. Moreover, public timeline expectations can influence technology adoption, investment decisions, and educational choices, making them important factors in shaping AI's actual development trajectory.

Third, we examine whether beliefs about if and when AI-driven societal change will occur are linked to self-reported AI knowledge and demographic factors. Recent studies have demonstrated that general knowledge about AI remains low across various populations (e.g., Almaraz-López et al., 2023; Rehman et al., 2024; Yim & Wegerif, 2024). We do not expect AI knowledge to predict whether different scenarios will occur, as the Theory of Planned Behavior has shown that the relationship between knowledge and attitudes/behavior is complex and non-linear (e.g., Ajzen et al., 2011). This lack of correlation has also been observed in AI-specific contexts; for instance, one study found no relationship between physicians' attitudes toward AI in medicine and their self-reported AI knowledge (Al-Medfa et al., 2023). However, AI knowledge might correlate with predictions about when AI-driven societal transformation will occur, as temporal predictions should be more knowledge-dependent than predictions about whether changes will happen.

Given demographic patterns observed in previous research, we also explore how gender, age, and education influence AI timeline expectations. First, we would expect gender to play a significant role in shaping expectations about AI development timelines. Men may predict more accelerated timelines compared to women, consistent with their generally higher levels of AI optimism and support for development (Zhang & Dafoe, 2019; Grassini & Sævild Ree, 2023). However, this gender difference might be more pronounced for perceptions of beneficial AI developments rather than potential risks, as previous research found significant gender differences in "AI Hope" but not "AI Doom" measures (Grassini & Sævild Ree, 2023). For education, we would anticipate that individuals with higher education levels will predict earlier timelines for AI advancement. College graduates and those with technical backgrounds have consistently shown greater support for AI development (Zhang & Dafoe, 2019), suggesting they may see technical challenges as easier to overcome. Regarding age, the expectations are less clear-cut. While Liang and Lee (2017) found older individuals report higher levels of fear toward autonomous robots and AI, Grassini & Sævild Ree (2023) found that age did not significantly impact either AI optimism or pessimism. This suggests that age might influence the perceived desirability of AI advances without necessarily affecting timeline expectations.

Thus, this study addresses gaps in previous literature by examining if and when the Swedish public expects major AI-driven societal changes to occur, while also assessing if AI knowledge and demographic factors are associated with these expectations. Sweden provides a particularly interesting context for studying AI expectations, as Swedish citizens are consistently among the earliest adopters of new technologies in Europe (Jarzębowski et al. 2024). This technological progressiveness suggests that Swedes may have formed more concrete opinions about AI's trajectory than citizens in many other countries, making their expectations particularly valuable to study. We investigate temporal expectations across six key dimensions of potential AI impact: unemployment, democratic functioning, living standards, medical advances, general workplace automation, and the development of unaligned superintelligent systems. The scenarios can be categorized into three broader groupings: positive developments, negative consequences, and general technological milestones.

In the positive domain, we investigate expectations regarding medical advances and improvements in living standards. The medical scenario builds upon recent breakthrough developments like AlphaFold, exploring public expectations about AI's potential to revolutionize healthcare through enhanced drug discovery, personalized medicine, and diagnostic capabilities (Thornton et al., 2021; Johnson et al., 2021). Similarly, the living standards scenario is grounded on expectations about AI's potential to boost economic productivity, optimize resource allocation, and reduce costs of goods and services through automated processes and enhanced efficiency (Davidson, 2021).

The negative scenarios focus on potential societal challenges: unemployment and threats to democracy. The unemployment scenario addresses concerns about AI-driven automation potentially displacing jobs, possibly at a pace exceeding the creation of new employment opportunities (Moser, 2022). The democratic functioning scenario is based upon AI's potential to undermine democratic processes through advanced misinformation capabilities, manipulation of public discourse, and enhanced surveillance technologies (Jungherr, 2023).

Two additional scenarios examine more general technological developments, distinguished by their more neutral or ambiguous nature in terms of immediate societal impact. The first concerns the development of AGI, a milestone that, while transformative, cannot be easily categorized as purely positive or negative. Like previous technological revolutions such as electricity or the internet, AGI's impact would likely be multifaceted and complex and depend on how closely the AGI-technology is aligned with human interests or not (Russell, 2019). The second general scenario examines expectations regarding the emergence of misaligned superintelligent systems—a development that, while generally considered negative due to its existential risk implications, represents a distinct category of concern from more immediate societal impacts. Such systems could include autonomous decision-making AI that prioritizes its own objectives over human well-being, self-replicating AI optimizing resource use in ways that deplete essential supplies, or strategic AI systems that manipulate information or social structures to entrench their control.

Using the above potential scenarios related to AI-developments, we examine Swedish public opinion regarding the likelihood and expected timing of these events. By examining both near to medium-term societal challenges (such as labor market changes and democratic challenges) and fundamental technological breakthroughs (such as AGI and superintelligence), we can build a more comprehensive understanding of how the public conceptualizes AI's future trajectory - from concrete societal impacts to more complex and potentially existential developments.

**Methods**
*Participants*
A random sample of 4,046 individuals aged 18 and over in the Swedish population were invited to participate in the study. Participants were first given the opportunity to complete the survey online. After several reminders, those who had not yet responded were mailed a paper version of the survey along with a reply envelope. This mixed-mode approach was employed to maximize response rates. The survey was conducted between June 24 and October 21, 2024, resulting in 1,026 completed responses (response rate: 25.4%).

*Measures*
The survey included several questions AI and its potential societal impacts. First, respondents were asked to rate their self-assessed knowledge about AI development and its potential societal effects on a 4-point scale ranging from "No knowledge at all" to "Very good knowledge." The survey then measured expectations across six key domains using binary response options (will/will not lead to) followed by timing estimates for those who answered affirmatively. These domains included: unemployment (whether AI would lead to a major increase in unemployment), democracy (whether AI would lead to a major deterioration of democracy), standard of living (whether AI would lead to a major improvement in people's standard of living), medical advances (whether AI would lead to major medical breakthroughs), automation (whether AI would lead to computers/robots that could perform all types of jobs as well as humans), and superintelligence (whether AI would lead to superintelligent machines beyond human control). For respondents who indicated they believed these developments would occur, follow-up questions assessed expected timing using a 6-point scale ranging from "Less than 1 year" to "More than 20 years.". The complete set of survey questions regarding AI expectations and their timelines, translated from Swedish to English, is provided in Appendix A.

Age was measured as a continuous variable (birth year). Gender was assessed as a categorical variable with response options "Female," "Male," and "Other." Educational attainment was measured using a six-point scale ranging from "Not completed primary school" to "Post-secondary education".

*Analysis*
Our analysis proceeded in two main stages. First, we conducted descriptive analyses to examine the proportion of respondents who believed each AI development would occur, calculating percentages and frequencies for each scenario.

Second, to identify patterns in how individuals viewed different AI developments, we employed latent class analysis (LCA) using the poLCA package in R. LCA was chosen as it allows for the identification of underlying subgroups with distinct response patterns across multiple binary items. We fitted models with two to five latent classes, comparing their fit using the Bayesian Information Criterion (BIC) and Akaike Information Criterion (AIC). For model selection, we prioritized BIC while also considering AIC and theoretical interpretability, while also demanding that the model fits the data on account of $L^2$ and TBVR to ensure stable solutions and avoid local maxima, each model was estimated with 20 random starts and a maximum of 3,000 iterations.

For the temporal analysis, we included two groups of respondents: those who believed the development would not occur (answering "will not lead to") and those who provided a timeline estimate for when they believed the development would occur. For those who believed the development would occur, we analyzed the distribution of expected timelines across six time periods (from "less than 1 year" to "more than 20 years"). The percentages for both the "Never" responses and the timeline estimates were calculated using the total number of respondents who either answered "will not lead to" or provided a timeline estimate as the base. This analytical approach was chosen for two main reasons: First, to provide a complete picture of the temporal expectations in the population, where "Never" represents a legitimate temporal perspective rather than a missing value. Second, because the proportion of respondents who believed each scenario would occur varied substantially across scenarios, using only those who believed in each development as the base would have made comparisons between scenarios misleading.

For analysis purposes, demographic variables were recoded to create meaningful categories. Age was categorized into three groups based on tertiles of the age distribution: "Young" (18-44 years), "Middle" (45-64 years), and "Old" (65-85 years). Gender was analyzed as a binary variable ("Female" and "Male"), with the three respondents who selected "Other" excluded from this analysis due to small cell counts. Educational attainment was recoded into a binary variable distinguishing between "Low/Medium education" (comprising those who had not completed primary school through those who had completed three years of secondary education) and "High education" (comprising those with at least some post-secondary education). Self-assessed AI knowledge was similarly dichotomized into "Low/Medium AI knowledge" and "High AI knowledge" categories, with the former combining respondents reporting "No knowledge at all" or "Little knowledge" and the latter combining those reporting "Good knowledge" or "Very good knowledge."

The LCA was conducted on five binary items measuring beliefs about whether AI would lead to: increased unemployment, deterioration of democracy, improved standard of living, medical advances, and uncontrollable superintelligent machines. The question about AI achieving human-level performance across all tasks was excluded from this analysis since it represents a technological milestone that is value-neutral in terms of positive or negative societal impact, unlike the other five scenarios which have clearer positive or negative implications. To handle missing data, we employed complete case analysis, where cases with missing values on any of the five items were excluded from the LCA. All analyses were conducted using R version 4.3.1 with the latent class analysis performed using the poLCA package.

To examine AI-knowledge and the socio-demographic predictors of latent class membership, we conducted a multinomial logistic regression using the multinom function from the nnet package (Venables & Ripley, 2002) in R. The dependent variable was the most likely class membership assigned to each respondent based on the LCA results. Independent variables included education level (categorized as high or low), gender, age (categorized as young, middle, or old), and self-reported AI knowledge level (categorized as high or low). Odds ratios (OR) with 95% confidence intervals were calculated to measure the strength and direction of associations. Where resulting OR were <1, reciprocal ORs (1/OR) for ease of interpretation.

**Results**
*Expected Occurrence of Transformative AI Developments*
The survey revealed varying levels of belief in different transformative AI developments. Medical breakthroughs emerged as the most widely anticipated outcome, with 82.6% of respondents believing this will occur at some point. This was followed by considerably lower expectations for other developments. 40.9% of respondents believed AI would lead to mass unemployment, while 40.3% anticipated improved living standards. Concerns about democracy harm were expressed by 38.7% of respondents, and 33.9% believed uncontrollable superintelligent AI would emerge. The development of human-level AI was seen as the least likely scenario, with only 28.4% of respondents believing this would occur.

The optimal number of latent classes was determined by comparing models with two to five classes using multiple fit indices. The Bayesian Information Criterion (BIC) suggested a three-class solution (BIC = 5493.568) as optimal, while the Akaike Information Criterion (AIC) favored a four-class solution (AIC = 5410.406). The three-cluster model yielded a non-significant difference between observed and expected cell frequencies on $L^2$ and $X^2$ statistics (p=.13 and .11 respectively) which is not the case for the four-cluster model (p<.01) and a reasonable TBVR of 4.14. One pairwise residual correlation exceeds 1.96 with all other correlations being <.7 on five variables, which indicates an acceptable model (Masyn, 2013). Given the principle of parsimony and the stronger theoretical interpretability, the three-class solution was selected for further analysis.

The latent class analysis revealed three distinct groups with varying attitudes towards artificial intelligence and its potential societal impacts. The largest group, comprising 46.65% of respondents (Class 2), displayed predominantly optimistic views about AI development. This group showed particularly high confidence in AI's potential for medical advancements (96.46% probability) and improvements in living standards (60.11% probability), while expressing small concern about potential negative consequences such as mass unemployment (21.25%), democratic deterioration (13.72%), or uncontrollable artificial superintelligence (10.96%).

The second largest group (Class 3), representing 42.20% of respondents, exhibited a more complex and ambivalent attitude toward AI. While recognizing its potential benefits, particularly in medicine (80.74% probability), this group showed substantial concern about potential risks. A majority within this class anticipated negative outcomes such as mass unemployment (70.17%), democratic deterioration (69.18%), and uncontrollable artificial superintelligence (65.24%). Only 30.33% of this group expected improvements in living standards.

The smallest group (Class 1), comprising 11.16% of respondents, demonstrated a distinctly skeptical or indifferent attitude toward AI's potential impact. This group consistently showed low probabilities for both positive and negative outcomes. Notably, none of the respondents in this class believed AI would improve living standards (0% probability), and only 33.29% expected medical advances. Similarly, they showed low probabilities for negative outcomes such as mass unemployment (7.84%) or democratic deterioration (28.26%). These findings suggest a polarized perspective on AI's societal impact, with the majority of respondents split between optimistic and concerned viewpoints, while a smaller group remains skeptical about AI's transformative potential in either direction.

The multinomial logistic regression analysis revealed several predictors of AI attitude class membership, with the optimistic class (Class 2) serving as the reference category (see figure 1). Education level emerged as the strongest predictor (p < .001), with higher education being associated with 2.56 times higher odds of belonging to the optimistic class compared to the skeptical class, and 1.69 times higher odds compared to the ambivalent class.
Gender showed a significant effect (p = .015), with men being 1.49 times more likely to belong to the ambivalent class compared to the optimistic class. However, gender did not significantly differentiate between optimistic and skeptical class membership (OR = 0.91). Finally, self-reported AI knowledge was also a significant predictor (p = .007). Individuals with high AI knowledge were 2.78 times more likely to belong to the optimistic class compared to the skeptical class, while the difference between optimistic and ambivalent class membership was not significant (OR = 1.25). No overall significant effect was found between age and class membership (p = 0.065), although the odds ratio for the difference between the optimistic and ambivalent class did not overlap 1 (OR = 1.64), with older respondents favoring the optimistic class compared to younger.

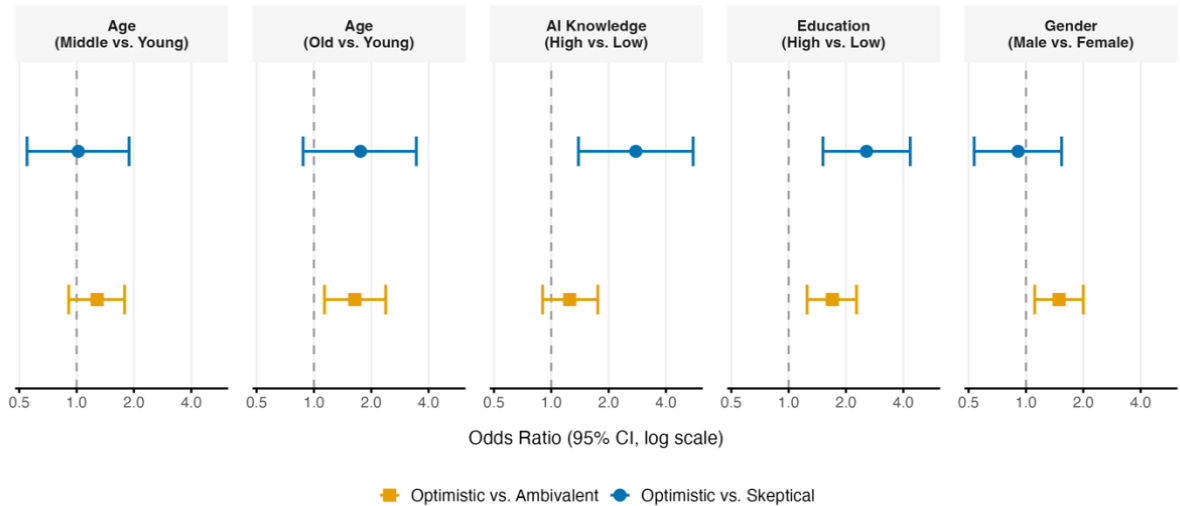

Figure 1: Predictors of AI Class Membership

Note: OR > 1 indicates higher odds for the first category in each pair.

*Timelines among respondents who believe transformative AI-developments will occur*
Among those who believed AI-developments would occur, the expected timelines varied significantly across different scenarios (see Table 1). Medical breakthroughs were not only the most widely expected (83% believing it will occur), but also anticipated relatively soon, with 35% of respondents expecting this within 6-10 years, and another 21% expecting it within 1-5 years. Mass unemployment and improved living standards showed similar overall belief rates (around 40%), though with different timeline distributions - unemployment was expected sooner, with 19% anticipating it in 6-10 years, while improved living standards showed a more gradual distribution across time periods.

Table 1: Percentage of respondents expecting each development by time period

| Scenario | N | < 1 year | 1-5 years | 6-10 years | 11-15 years | 16-20 years | > 20 years | Never |
|---|---|---|---|---|---|---|---|---|
| Major increase in unemployment | 949 | 0.6 | 8.4 | 18.8 | 7.9 | 3.1 | 2.1 | 59.1 |
| Major deterioration of democracy | 938 | 1.6 | 11.3 | 14.9 | 6.9 | 2.6 | 1.4 | 61.3 |
| Major improvement in standard of living | 944 | 0.3 | 6.2 | 13.7 | 11.8 | 4.6 | 3.7 | 59.7 |
| Major medical breakthroughs | 943 | 1.2 | 20.9 | 35.2 | 15.2 | 6 | 4.1 | 17.4 |
| AI performing all jobs as well as humans | 938 | 0.7 | 4.1 | 8.3 | 6.7 | 3.1 | 5.4 | 71.6 |
| Uncontrollable Superintelligence | 937 | 0.3 | 2.8 | 6.4 | 7.6 | 6.1 | 10.8 | 66.1 |

Democracy deterioration was expected by 39% of respondents, with the highest concentration (15%) expecting it in the 6-10 year range, followed by 11% anticipating it within 1-5 years. The more dramatic technological developments showed both higher skepticism and longer expected timelines. For uncontrollable superintelligent AI, 66% believed it would never happen, while 11% expected it beyond 20 years - the highest long-term percentage among all scenarios. Similarly, 72% doubted that computers/robots would reach human-level performance in all jobs, with believers showing a relatively even distribution across different time periods. This pattern suggests that respondents generally expect more dramatic technological developments to take longer or not happen at all, indicating a measured understanding of the relationship between different scenarios.

Analysis of correlations between demographic variables and AI timeline predictions (see Table 2) revealed that education showed no significant correlations with any of the AI timeline estimates (all rs < |.09|), and gender similarly displayed no significant relationships with timeline predictions (all rs < |.07|). Age showed two significant negative correlations: with unemployment predictions (rs = -.13, p < .05) and superintelligence predictions (rs = -.19, p < .01), suggesting older respondents tended to predict shorter timelines for these developments. AI Knowledge had only one significant correlation: a weak positive relationship with superintelligence estimates (rs = .14, p < .05), indicating that those with higher self-rated AI knowledge tended to predict longer timelines for uncontrollable superintelligence.

Table 2: Correlations Between Demographics, AI-knowledge and AI Timeline Estimates

| Variables | 1 | 2 | 3 | 4 | 5 | 6 | 7 | 8 | 9 | 10 |
|---|---|---|---|---|---|---|---|---|---|---|
| 1. EDU | | -0.111** | -0.124** | 0.186** | 0.085 | 0.024 | -0.031 | -0.015 | 0.083 | 0.042 |
| 2. GEN | | | -0.023 | 0.152** | 0.063 | -0.073 | 0.033 | 0.030 | 0.059 | 0.032 |
| 3. AGE | | | | -0.314** | -0.127* | -0.070 | -0.037 | -0.069 | -0.119 | -0.190** |
| 4. AIK | | | | | -0.024 | -0.049 | -0.026 | -0.062 | 0.019 | 0.144* |
| 5. UNE | | | | | | 0.594** | 0.587** | 0.368** | 0.413** | 0.509** |
| 6. DEM | | | | | | | 0.369** | 0.405** | 0.444** | 0.562** |
| 7. IMP | | | | | | | | 0.591** | 0.567** | 0.525** |
| 8. MED | | | | | | | | | 0.425** | 0.419** |
| 9. ROB | | | | | | | | | | 0.701** |
| 10. SUP | | | | | | | | | | |

Note: *p < 0.05, **p < 0.01. EDU = Education level; GEN = Gender (1=Female, 2=Male); AGE = Age; AIK = AI Knowledge; UNE = Major increase in unemployment; DEM = Major deterioration of democracy; IMP = Major improvement in standard of living; MED = Major medical breakthroughs; ROB = AI performing all jobs as well as humans; SUP = Superintelligent machines beyond human control.

**General discussion**

This study sought to fill a gap in the understanding of public expectations regarding the timeline of significant AI-driven societal changes, focusing on the Swedish population as an early adopter of new technologies. The findings reveal diverse public views, showing both optimism and concern about the potential impacts of artificial intelligence. Notably, the majority of respondents anticipate that AI will lead to major medical breakthroughs, likely reflecting a widespread belief in the positive potential of AI to advance healthcare. However, there is a significant divergence in expectations concerning other domains such as mass unemployment, severe deterioration of democratic systems, substantially improved living standards, and the emergence of artificial general intelligence and uncontrollable superintelligent systems. In fact, most Swedes believe that none of the latter five scenarios will ever occur.

The high expectation of medical advancements within a relatively short timeframe may underscore the public's recognition of recent AI achievements in healthcare, such as diagnostic improvements and personalized medicine. The anticipation of medical breakthroughs within the next 6–10 years suggests that the public expects rapid progress in this field. The data collection period partly overlapped with the researchers behind AlphaFold receiving the 2024 Nobel Prize in Chemistry, an event that garnered substantial public attention.

Conversely, expectations about AI leading to mass unemployment, deterioration of democracy, and the development of uncontrollable superintelligent machines are less common but still prevalent as approximately one-third of respondents expressed these concerns. The timelines for these developments are generally projected over a longer horizon, particularly for uncontrollable superintelligent AI and human-level AI, which many respondents believe will not occur for more than 20 years. This temporal distancing may reflect a perception that such profound changes require more time to unfold or a degree of skepticism about the feasibility of these outcomes.

In the expert literature on AI timelines as well as among superforecasters, the expected arrival of artificial general intelligence (AGI) has gained much attention. Recent large-scale surveys provide insights into expert predictions: the 2023 AI Impacts survey, which included 2,778 AI researchers who had published at top-tier AI venues, indicated that experts collectively estimated a 10% probability of high-level machine intelligence (HLMI) arriving by 2027 and a 50% probability by 2047, where HLMI was defined as machines capable of outperforming humans in all tasks both in terms of effectiveness and cost-efficiency (Grace et al. 2024). Our survey question asking "whether AI would lead to computers/robots that could perform all types of jobs as well as humans" closely resembles what Grace et al. called the Full Automation of Labor (FAOL) framing - defined as when "all occupations are fully automatable". For this occupation-focused framing, experts predicted a much later timeline (50% probability by 2116) compared to the task-focused HLMI framing. This large difference based solely on framing suggests that how we posed our question may have influenced our respondents toward greater skepticism. While most of our respondents did not project AGI to ever occur, 12% anticipated it to happen within 10 years. This variation in expectations within the public sample mirrors the wide range of predictions among experts documented by Grace et al., where the 50% probability estimates for occupation-based automation extended into the next century.

The latent class analysis provides further insight into the heterogeneity of public opinion. The largest group, representing nearly half of the respondents, is predominantly optimistic about AI's benefits and exhibits modest concern about its potential risks. The second largest group, however, expresses a more ambivalent stance, recognizing some potential benefits (medical advances) but significant risks associated with AI in the form of mass unemployment, democratic deterioration as well as uncontrollable artificial superintelligence. Their substantial concern about negative outcomes echoes the warnings of experts who emphasize the existential risks and ethical challenges posed by powerful AI-systems (Bostrom, 2017; Hinton, as cited in Schechner & Seetharaman, 2023). The smallest group is characterized by skepticism or indifference toward AI's transformative potential, whether positive or negative. This may indicate a segment of the population that is either less informed about AI developments or less convinced of their significance.

The latent class analysis revealed significant demographic differences between the classes. We did not find the optimistic gender effect of men identified in previous studies (eg. Grassini & Sævild Ree, 2023), Swedish men were more ambivalent than optimistic, compared to women, and gender did not significantly differentiate between skeptical and optimistic class membership. Instead, our analysis identified education level as the strongest predictor of class membership, with higher education significantly associated with belonging to the optimistic class rather than the skeptical or ambivalent classes. The substantial effect for education is noteworthy, and the question is what underlying attitudes form these results. A possibility is that individuals with lower levels of education are less confident that the benefits of AI will be equitably distributed across society and may question whether such developments will lead to meaningful improvements in their own lives, given existing social and economic disparities. Self-reported AI-knowledge decreased the likelihood of belonging to the skeptical class. Age did not show any significant association with AI positivism or scepticism, in line with previous findings (Grassini & Sævild Ree, 2023).

Regarding specific timeline predictions, our correlation analysis between demographic variables and AI timeline estimates revealed interesting patterns. Education and gender showed no significant correlations with any AI timeline estimates, contradicting expectations that these factors would influence timeline perceptions. Age, however, showed significant negative correlations with both unemployment and superintelligence predictions, suggesting older respondents tended to predict shorter timelines for these developments. AI knowledge showed a weak positive relationship with superintelligence estimates, indicating that those with higher self-rated AI knowledge tended to predict longer timelines for uncontrollable superintelligence. These findings partially align with our expectations discussed in the introduction—while education and gender strongly predict overall AI attitudes as anticipated, they surprisingly show minimal influence on specific timeline predictions. Additionally, the relationship between AI knowledge and timeline expectations revealed a more nuanced pattern than initially expected, with expertise correlating with more conservative rather than accelerated timeline estimates for advanced AI capabilities. These results add important nuance to our understanding of public expectations, suggesting that while demographic factors do influence AI timeline predictions, their effects are subtle and specific rather than broad and systematic.

In relation to policy discussions about future AI developments, our findings about the public's expectations do not align with the probability assessments implied by AI catastrophists or with those suggested by AI progressives. Instead, we find a more nuanced picture where the Swedish public largely considers progress in specific domains like medicine to be probable, while judging more dramatic societal transformations as unlikely. The temporal variation in when different AI developments are expected to occur may highlight the need for policy frameworks that can address both near-term applications and potential longer-term challenges. The widespread belief that medical AI applications will materialize, coupled with more divided predictions about societal risks, suggests that the public may be most receptive to governance frameworks that address concrete, domain-specific developments (like medicine) rather than broader existential concerns. Yet policymakers must also consider whether the public's assessment that transformative AI developments are unlikely might lead to underinvestment in long-term safety measures and governance structures. The study also underscores the heterogeneity in public probability assessments, with distinct groups holding markedly different views about what developments are likely to occur in AI's trajectory. This diversity of predictions about future AI developments suggests that any policy approach will need to engage with and address multiple, sometimes conflicting, public expectations rather than assuming a consensus view.

Several limitations of this study should be acknowledged. The response rate of 25% could introduce non-response bias, as participants who completed the survey might differ systematically from those who declined to participate, particularly in their interest in or knowledge about AI technologies. Additionally, while the study focused on six key scenarios related to transformative AI development, several other potentially important developments were not included, such as AI's impact on global security and warfare, environmental sustainability, education systems, social relationships and communication patterns, or the potential emergence of human-AI hybrid societies (Bostrom, 2017; Russell 2019). It is also unclear how informed these guesses about AI are in the general public, and work needs to be done in the future regarding the reliability of these guesses as well as the survey measures used to capture expected occurrences and timelines of AI development. This includes framing effects that may influence how respondents interpret and answer questions about future AI capabilities, as shown by recent expert surveys where different framings of similar AI milestones led to substantially different timeline predictions (Grace et al., 2024).

In conclusion, this study provides insights into the Swedish public's expectations about AI-driven societal changes, revealing a spectrum of optimism and concern. As AI continues to advance, understanding and engaging with public expectations will be essential in guiding its development in ways that align with societal values and interests.

**Acknowledgments**
The authors would like to thank Sebastian Lundmark and Felix Cassel for their help with data collection and valuable feedback on the survey questions. Language editing and R code development were supported by Claude, a large language model developed by Anthropic.

**Appendix A**

BLOCK: AI (INTRO + INITIAL QUESTION)
First, some questions about your views on how artificial intelligence (AI) may affect society in the future.
How much knowledge do you consider yourself to have about the development of AI (artificial intelligence) and its possible impact on society?
- Very good knowledge (1)
- Good knowledge (2)
- Little knowledge (3)
- No knowledge at all (4)

BLOCK: AI (UNEMPLOYMENT)
Do you think AI development will lead to a very large increase in unemployment in the future or not?
- Will lead to a very large increase in unemployment (1)
- Will not lead to a very large increase in unemployment (2)

[If respondent selected "Will lead to a very large increase in unemployment":] When do you think AI development will lead to a very large increase in unemployment?
- In less than 1 year (1)
- In 1-5 years (2)
- In 6-10 years (3)
- In 11-15 years (4)
- In 16-20 years (5)
- In more than 20 years (6)

BLOCK: AI (DETERIORATED DEMOCRACY)
Do you think AI development will lead to a very large deterioration of democracy in the future or not?
- Will lead to a very large deterioration of democracy (1)
- Will not lead to a very large deterioration of democracy (2)

[If respondent selected "Will lead to a very large deterioration of democracy":] When do you think AI development will lead to a very large deterioration of democracy?
- In less than 1 year (1)
- In 1-5 years (2)
- In 6-10 years (3)
- In 11-15 years (4)
- In 16-20 years (5)
- In more than 20 years (6)

BLOCK: AI (STANDARD OF LIVING)
Do you think AI development will lead to a very large improvement in people's standard of living in the future or not?
- Will lead to a very large improvement in people's standard of living (1)
- Will not lead to a very large improvement in people's standard of living (2)

[If respondent selected "Will lead to a very large improvement in people's standard of living":] When do you think AI development will lead to a very large improvement in people's standard of living?
- In less than 1 year (1)
- In 1-5 years (2)

- In 6-10 years (3)
- In 11-15 years (4)
- In 16-20 years (5)
- In more than 20 years (6)

BLOCK: AI (MEDICAL ADVANCES)

Do you think AI development will lead to very significant medical advances in the future or not?
- Will lead to very significant medical advances (1)
- Will not lead to very significant medical advances (2)

[If respondent selected "Will lead to very significant medical advances":] When do you think AI development will lead to very significant medical advances?
- In less than 1 year (1)
- In 1-5 years (2)
- In 6-10 years (3)
- In 11-15 years (4)
- In 16-20 years (5)
- In more than 20 years (6)

BLOCK: AI (ROBOTS)

Do you think AI development will lead to computers or robots that can perform all types of jobs as well as humans in the future or not?
- Will lead to computers or robots that can perform all types of jobs as well as humans (1)
- Will not lead to computers or robots that can perform all types of jobs as well as humans (2)

[If respondent selected "Will lead to computers or robots that can perform all types of jobs as well as humans":] When do you think AI development will lead to computers or robots that can perform all types of jobs as well as humans?
- In less than 1 year (1)
- In 1-5 years (2)
- In 6-10 years (3)
- In 11-15 years (4)
- In 16-20 years (5)
- In more than 20 years (6)

BLOCK: AI (SUPERINTELLIGENCE)

Do you think AI development will lead to superintelligent machines that become impossible for humans to control in the future or not?
- Will lead to superintelligent machines that become impossible for humans to control (1)
- Will not lead to superintelligent machines that become impossible for humans to control (2)

[If respondent selected "Will lead to superintelligent machines that become impossible for humans to control":] When do you think AI development will lead to superintelligent machines that become impossible for humans to control?
- In less than 1 year (1)

- In 1-5 years (2)
- In 6-10 years (3)
- In 11-15 years (4)
- In 16-20 years (5)
- In more than 20 years (6)